\documentclass[aps,prl,twocolumn,superscriptaddress,floatfid,nofootinbib]{revtex4-2}

\usepackage{amsmath,amssymb}      
\usepackage{graphicx}             
\usepackage[colorlinks=true,       
            linkcolor=blue,
            citecolor=blue,
            urlcolor=blue]{hyperref}
\usepackage{booktabs}   
\usepackage[caption=false]{subfig}
                     
\graphicspath{{paper_figs/}}

\usepackage{float}               
\usepackage{mathtools}
\usepackage{dutchcal}

\newcommand{\tr}{\text{Tr}}
\newcommand{\sinhc}{\text{sinhc}}

\usepackage{pdfpages}
\usepackage{pgffor}
\makeatletter
\AtBeginDocument{\let\LS@rot\@undefined}
\makeatother
\def\supplementfilename{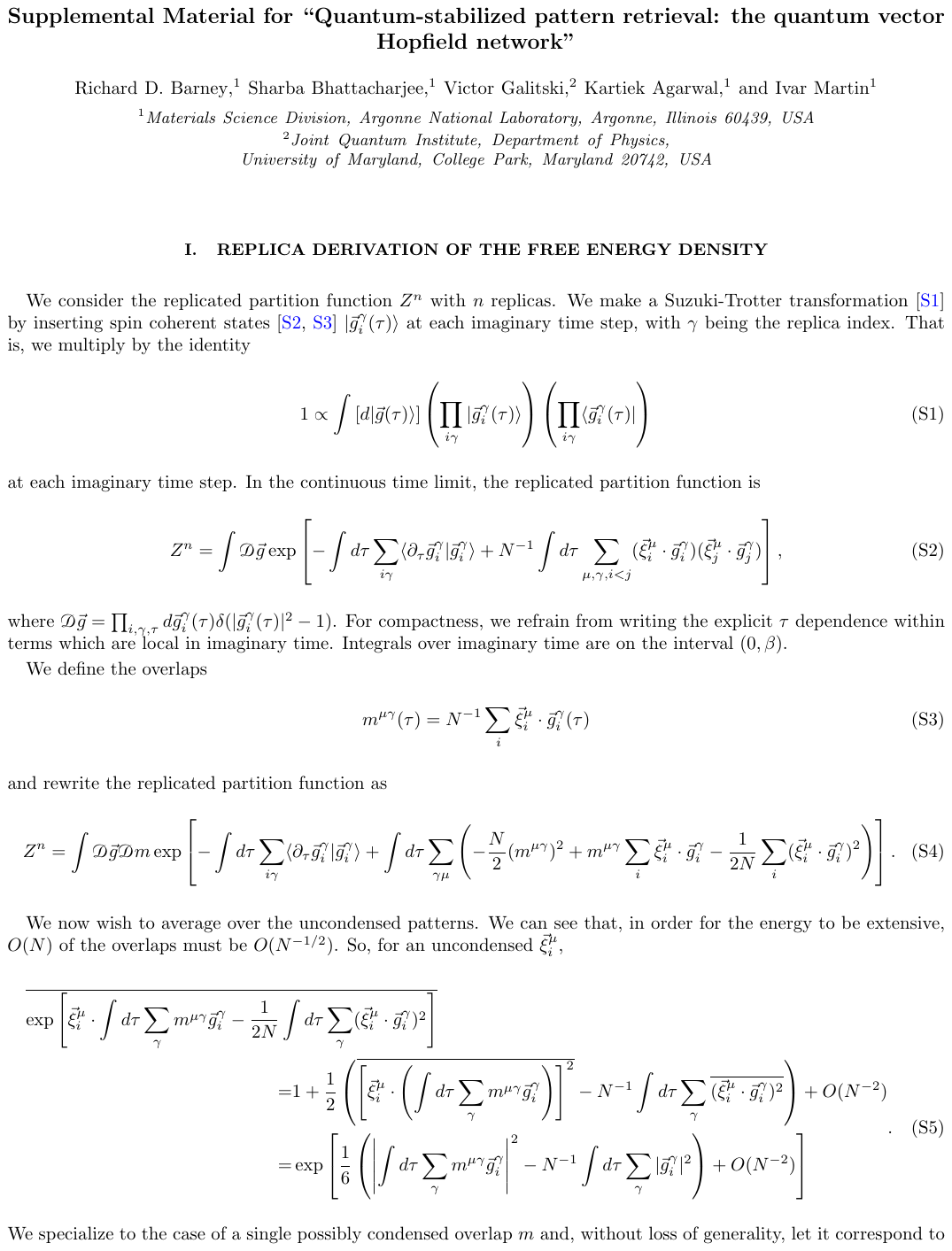}
\pdfximage{\supplementfilename}
\edef\numsupppages{\the\pdflastximagepages}

\begin{document}

\title{Quantum-stabilized patterns in a vector Hopfield network
}

\author{Richard D. Barney}
\affiliation{Materials Science Division, Argonne National Laboratory, Argonne, Illinois 60439, USA}

\author{Sharba Bhattacharjee}
\affiliation{Materials Science Division, Argonne National Laboratory, Argonne, Illinois 60439, USA}

\author{Victor Galitski}
\affiliation{Joint Quantum Institute,
Department of Physics, University of Maryland, College Park, Maryland 20742, USA}

\author{Kartiek Agarwal}
\affiliation{Materials Science Division, Argonne National Laboratory, Argonne, Illinois 60439, USA}

\author{Ivar Martin}
\affiliation{Materials Science Division, Argonne National Laboratory, Argonne, Illinois 60439, USA}

\begin{abstract}
    We introduce the quantum vector Hopfield network, in which patterns are formed by orientations of quantum vector spins;  quantum dynamics arise intrinsically from the non-commutativity of the spin operators. We derive the equations of state and the phase diagrams for this network as well as its classical counterpart. We find that quantum fluctuations, surprisingly, stabilize the stored patterns. Both the critical retrieval temperature and the target pattern overlap are enhanced relative to the classical network. Additionally, we find that this enhancement grows with pattern loading up to network capacity. We interpret this effect as an analog of quantum order-by-disorder, a mechanism by which quantum fluctuations promote the formation of ordered phases. These findings offer a new route to quantum-enhanced associative memory.
\end{abstract}

\maketitle

\paragraph{\textbf{Introduction:}}The Hopfield network~\cite{Hopfield1982} was originally introduced as a model of biological neural networks, demonstrating that associative memory can emerge from attractor dynamics in a network of all-to-all connected binary neurons with  couplings encoding memory patterns (``Hebbian learning''~\cite{Hebb1949}). In recent years, there has been a resurgence of interest in such networks. Generalizations of the original Hopfield network to higher-order interactions, referred to as modern Hopfield networks, are able to encode a number of patterns which grows as higher order polynomials or even exponentially with system size, well beyond the linear scaling of the original network~\cite{Gardner1987-wq,Krotov2018-gf, krotov2016dense,Demircigil2017-as,Lucibello2024-bv}. Apart from their role as models of associative memory, a fundamental connection has been pointed out between Hopfield networks and certain deep-learning architectures~\cite{Krotov2023-gn,Ambrogioni2024-sb}, including the attention mechanism within transformers~\cite{Ramsauer2020-ks}, which have enjoyed success in a variety of machine learning tasks. The methods developed for Hopfield networks~\cite{Amit1987-nv,Gardner1988-eu,amit1989modeling} can thus be brought to bear on modern machine learning architectures. Hopfield networks are themselves prototypical energy-based models, a framework recently argued to offer advantages over standard likelihood-based architectures~\cite{Dawid2024-fw}.

Quantum generalizations of associative memory~\cite{Ventura2000-jb,Lewenstein2021-xo,Rebentrost2018-bi,Amin2018-dj,Fiorelli2020-uv,Marsh2025-py,Schuld2015-wp} have also attracted renewed attention as a setting in which the ideas of neural network theory, many-body physics, and quantum information converge. A key choice in designing a quantum Hopfield network is the way in which non-commuting dynamics are introduced. Several models accomplish this by adding an external transverse field, either in closed systems~\cite{Nishimori1996-dm,Ma1993-nq,Seki2015-op,Xie2024-ct} or open Lindbladian settings~\cite{Rotondo2018-oe,Fiorelli2022-oj,Bodeker2023-sk}. Because the field competes with the Hebbian interactions, its overall effect is analogous to temperature, suppressing the retrieval phase. A different quantization scheme with only $\sigma^y\sigma^z$  interactions ($\sigma^y$ and $\sigma^z$ being Pauli operators) was shown to have greater capacity than the original Hopfield network, but at the cost of pattern retrieval quality~\cite{Diamantini2006-ih}. This raises the question of whether alternative quantized Hopfield networks can be constructed which enhance retrieval without such trade-offs.

\begin{figure}
    \centering
    \subfloat{\includegraphics[width=\linewidth]{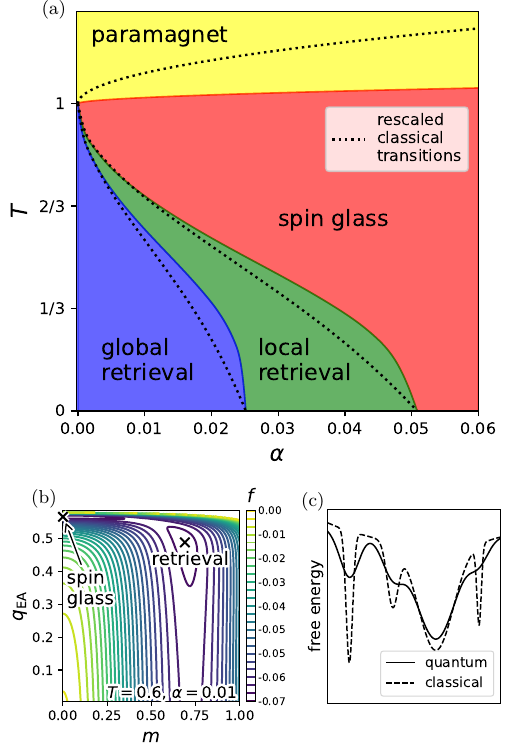}\label{fig:quant_PD}}
    \subfloat{\label{fig:basins}}
    \subfloat{\label{fig:smooth}}
    \caption{(a) The phase diagram of the QVHN. Memory states are retrievable up to a finite loading capacity of $\alpha_\text{r} N$ for $N$ spins, and $\alpha_\text{r} \approx 0.0508$. These states become global minima of the free energy landscape for $\alpha < \alpha_\text{gr} \approx 0.0252$; hence called the ``global retrieval" phase. The quantum network displays greater retrieval transition temperatures than the rescaled classical network (dotted lines).
    (b) The free energy landscape for the rescaled classical network at $T=0.6$, $\alpha=0.01$. Dependence on the parameter $r_d$ is removed by enforcing the saddle point constraint corresponding to Eq.~(\ref{eq:rd_class}). The retrieval minimum sits in a much broader valley than that of the spin glass minimum.
    (c) Schematic showing how quantum fluctuations raise the free energies of narrow classical valleys, such as those for spin glass states, more strongly than those of broad valleys corresponding to retrieval states. This in turn stabilizes the retrieval phase in the quantum model.}
\end{figure}

In this Letter we introduce and study such a network: the quantum vector Hopfield network (QVHN). This is a quantization of the three-dimensional classical vector Hopfield network (CVHN)~\cite{Nicoletti2025-pm,Gallavotti2025-fy}, which encodes patterns composed of continuous unit vectors instead of binary values. In contrast to the previously mentioned quantization schemes, we take an approach well-established in the field of quantum spin glasses~\cite{Bray1980-lk,Sachdev1993-bt,Ye1993-jn,Georges2001-yw} but, to our knowledge, not previously applied to associative memory, which is to promote the classical spin components to their corresponding quantum operators. Quantum dynamics then emerge intrinsically from the non-commutativity of the spin operators. We map the phase diagram for the QVHN (see Fig.~\ref{fig:quant_PD}) and demonstrate that it exhibits enhanced pattern  stability relative to the CVHN. In contrast to the transverse-field case, quantum fluctuations in the QVHN raise both the critical retrieval temperature and the overlap of the retrieved pattern with the target. We demonstrate that this enhancement goes beyond the baseline enhancement expected from the greater single-spin susceptibility of a quantum spin; it grows with pattern loading up to capacity. We interpret this enhancement as an analog of quantum order-by-disorder~\cite{Villain1980-vk,Shender1982,Henley1989-qj}, a phenomenon in which quantum fluctuations promote the formation of ordered phases.

\paragraph{\textbf{Model:}} A vector Hopfield network composed of $N$ spins, encoding $p$ patterns of the form $(\vec\xi_1^\mu,\dots,\vec\xi_N^\mu)$, where each $\vec\xi_i^\mu$ is a unit vector, is defined by the Hamiltonian
\begin{equation}
    H=-\frac 1 N\sum_{\mu=1}^p\sum_{i<j}(\vec\xi_i^\mu\cdot\vec\sigma_i)(\vec\xi_j^\mu\cdot\vec\sigma_j).
    \label{eq:model}
\end{equation}
We assume extensive pattern loading with $p\sim N$, $\alpha=p/N$ and let the $\vec\xi_i^\mu$ be uniformly and independently distributed on the unit sphere in three dimensions. For the QVHN, the $\vec\sigma_i$ are taken to be Pauli vectors, while the $\vec\sigma_i$ are classical Heisenberg spins in the CVHN, i.e., three-dimensional unit vectors.

The overlap of the network's state with the $\mu^\text{th}$ pattern is
\begin{equation}
    m^\mu=\frac 1 N\sum_i\vec\xi_i^\mu\cdot\vec\sigma_i.
    \label{eq:mattis}
\end{equation}
The $\mu^\text{th}$ Mattis magnetization~\cite{Mattis1976-sf} is then $\overline{\langle m^\mu\rangle}$,
where $\langle\cdots\rangle$ denotes the thermal average and $\overline{\cdots}$ denotes an average over disorder. We specialize to the case of a single possibly condensed Mattis magnetization $m=\overline{\langle m^1\rangle}$, which we can say, without loss of generality, corresponds with the first pattern.

Note that the normalization in Eq.~(\ref{eq:model}) realizes the same potential energy landscape in both the classical and quantum cases for classical/product states. An alternative choice is to normalize $H$ such that high temperature spin susceptibilities $\propto |\vec{\sigma}_i|^2$ are equal; this requires rescaling the classical spins to have norm $\sqrt{3}$ which boosts all energy scales, including transition temperatures by a factor of $3$. This rescaled CVHN yet exhibits poorer pattern retrievability and lower retrieval transition temperatures than the QVHN.


\paragraph{\textbf{Free energy density:}}
We have calculated the free energy density of the QVHN through the replica method~\cite{Edwards1975-ps,mezard1987spin}, under the replica symmetric ansatz and working within the static approximation, which averages the order parameters over imaginary time but respects individual spin fluctuations. This approximation captures the leading effects of quantum fluctuations and is commonly used in the analysis of quantum spin glasses and quantum Hopfield models~\cite{Bray1980-lk,Miller1993-ut,Goldschmidt1990-ru,Nishimori1996-dm}. Corrections to the approximation have been shown to be minute in a transverse-field quantum Hopfield network, due to the fact that the relevant non-zero Matsubara modes decay rapidly and feed back only weakly to the zeroth-mode susceptibility~\cite{Okajima2025-em}. In the Supplemental Material~\cite{Sup} we examine the effect of relaxing the approximation and find that the correction to the free energy density is parametrically small throughout the retrieval phase.

Within this framework, there are four relevant order parameters, in addition to $m$. The first two are the connected and disconnected parts of the imaginary-time-averaged spin autocorrelation
\begin{equation}
    q\hspace{-0.1em}=\hspace{-0.1em}\frac 1 N\hspace{-0.1em}\sum_i\hspace{-0.1em}\int\hspace{-0.1em}\frac{d\tau}{\beta}\overline{\langle\vec\sigma_i(\tau)\cdot\vec\sigma_i(0)\rangle}_c,\hspace{0.25em}q_\text{EA}\hspace{-0.1em}=\hspace{-0.1em}\frac 1 N\hspace{-0.1em}\sum_i\overline{|\langle\vec\sigma_i\rangle|^2},
\end{equation}
where $\langle\cdots\rangle_c$ denotes the connected part of the thermal average. The parameter $q$ is sensitive to the amount of quantum fluctuation of the spins. $q_\text{EA}$ is the Edwards-Anderson order parameter~\cite{Edwards1975-ps}, which measures the extent to which spins are $locally$ frozen (i.e., not thermally disordered). The last two parameters are the connected and disconnected parts of the imaginary-time-averaged autocorrelation of the uncondensed overlaps
\begin{equation}
    r\hspace{-0.1em}=\hspace{-0.1em}\frac 1\alpha\sum_{\mu>1}\int\frac{d\tau}{\beta}\overline{\langle m^\mu(\tau)m^\mu(0)\rangle_c},\hspace{0.7em}r_d\hspace{-0.1em}=\hspace{-0.1em}\frac 1\alpha\sum_{\mu>1}\overline{\langle m^\mu\rangle^2}.
\end{equation}
The parameters $q$ and $r$ contain the information about the quantum nature of the network. In the classical limit they are completely determined by $q_\text{EA}$ and $r_d$, respectively, and can therefore be substituted out. In particular, since $\vec\sigma_i(\tau)$ becomes constant in imaginary time in the classical limit, it follows that $q+q_\text{EA}=1$.

In terms of these parameters described above, the disorder-averaged free energy density is
\begin{align}
    &f=\begin{multlined}[t]
        \frac{\alpha}{2}\left(\frac 1\beta\ln(1-\beta q/3)-\frac{\beta q q_\text{EA}/9}{1-\beta q/3}+\frac{q}{3}\right)\\
        +\frac{\beta\alpha}6\left(r q+r q_\text{EA}+r_d q\right)+\frac{m^2}2+f_\text{ss},\label{eq:f}
    \end{multlined}\\
    &f_\text{ss}=-\frac 1\beta\int\mathcal d\vec h\ln\int\mathcal d\vec\ell~\tr e^{\beta\vec B\cdot\vec\sigma},\label{eq:fp_gint}
\end{align}
where $\mathcal d\vec h=dh_xdh_ydh_z(2\pi)^{-3/2}e^{-|\vec h|^2/2}$ and similarly for $\mathcal d\vec\ell$, and $f_\text{ss}$ is the effective single-site free energy. The fields $\vec h$ and $\vec\ell$ were introduced through Hubbard-Stratonovich transformations. The constant-in-imaginary-time single-site effective field is
\begin{equation}
    \vec B=m\vec\xi^1+\sqrt{\frac{\alpha r_d}3}\vec h+\sqrt{\frac{\alpha r}3}\vec\ell. \label{eq:eff_B}
\end{equation}
This effective field has two sources of noise. There is the classical term $\sqrt{\alpha r_d/3}\vec h$ which is analogous to the noise term which arises for the standard Hopfield network in Ref.~\cite{amit1989modeling}. There is also the term $\sqrt{\alpha r/3}\vec\ell$ which is sensitive to quantum fluctuations. From Eq.~(\ref{eq:fp_gint}) we see that $\vec\ell$ acts as annealed noise on the local field, being averaged inside the logarithm, while $\vec h$ acts as quenched disorder, being averaged outside the logarithm. That is, the quantum fluctuations do not simply add to the destabilizing effect of the classical noise. Details of the replica calculation of the free energy are found in the Supplemental Material~\cite{Sup}.

Note that $f_\text{ss}$ is independent of the orientation of $\vec\xi^1$. Simplifying Eq.~(\ref{eq:fp_gint}), we find
\begin{align}
    &f_\text{ss}=-\frac{\beta\alpha}{6}r-\frac 1 m\int\mathcal d\rho vw(v)\label{eq:fp_w}\\
    &w_\text{qu}(v)=\frac 1\beta\ln\left[\cosh(\beta v)+\frac{\beta^2 \alpha r}3\sinhc(\beta v)\right],\label{eq:w_qu}
\end{align}
where $\sinhc(x)=\sinh(x)/x$, $\mathcal d\rho=d\rho(2\pi)^{-1/2}e^{-\rho^2/2}$, and $v=m+\sqrt{\alpha r_d/3}~\rho$.

Equations~(\ref{eq:f}) and (\ref{eq:fp_w}) also hold for the CVHN, but with a different $w(v)$. Taking the quantum spin number to $\infty$ in Eq.~(\ref{eq:fp_gint}) and interpreting the trace as a sum over the degrees of freedom, yields an effective single-site free energy of the form found in Eq.~(\ref{eq:fp_w}), but with $w_\text{cl}(v)=\beta^{-1}\ln\sinhc(\beta v)$.

\paragraph{\textbf{Equations of state:}}
Using the above results, by variation of the free energy density with respect to the order parameters, we obtain the equations of state for the QVHN:
\begin{align}
    &m=\frac 1 m\int\mathcal d\rho\left(\sqrt{\frac 3{\alpha r_d}}\rho-\frac 1 m\right)vw_\text{qu}(v)\label{eq:q_eqs_m}\\
    &q=\frac{3}{\beta \alpha r_dm}\int\mathcal d\rho(\rho^2-1)vw_\text{qu}(v)\label{eq:q_eqs_qc}\\
    &q_\text{EA}=\frac 1 m\int\mathcal d\rho v(\partial_v w_\text{qu}(v))^2\label{eq:q_eqs_qd}\\
    &r=\frac{q/3}{1-\beta q/3},\qquad r_d=\frac{q_\text{EA}/3}{(1-\beta q/3)^2}.
\end{align}

The same can be done for the CVHN. A simplification arises from one of the saddle point equations, $q+q_\text{EA}=1$. This indicates that the CVHN has no quantum fluctuations, as must be the case. This allows us to eliminate the equations for $q$ and $r$, yielding the equations of state for the CVHN:
\begin{align}
    &m=\frac 1 m\int\mathcal d\rho\left(\sqrt{\frac 3{\alpha r_d}}-\frac 1 m\right)vw_\text{cl}(v)\label{eq:cl_eqs_m}\\
    &q_\text{EA}=1-\frac{3}{\beta m\alpha r_d}\int\mathcal d\rho(\rho^2-1)vw_\text{cl}(v)\label{eq:cl_eqs_qd}\\
    &r_d=\frac{q_\text{EA}/3}{[1-\beta(1-q_\text{EA})/3]^2}.\label{eq:rd_class}
\end{align}
Of course, in the classical case, there is no question of the applicability of the static treatment, so these equations are exact. These are equivalent to the equations of state obtained for the CVHN in Ref.~\cite{Nicoletti2025-pm}. An alternative calculation of the equations of state for the QVHN and CVHN through a mean-field method is presented in the Supplemental Material~\cite{Sup}.

Note that, as $\beta\rightarrow\infty$, $w(v)\rightarrow|v|$ for both the CVHN and QVHN. It follows that the networks have the same equations of state at zero temperature.
This is an outcome of the fact that the quantum ground state is a coherent state mimicking the classical ground state as closely as possible. It was found in Ref.~\cite{Nicoletti2025-pm} that the entire retrieval phase of the CVHN is replica symmetric. This is in contrast to the standard Hopfield network, which breaks replica symmetry at low temperatures~\cite{Amit1987-nv}. Since the QVHN's equations of state reduce to those of the CVHN at zero temperature, we conclude that the retrieval phase of the QVHN is also replica symmetric.

\paragraph{\textbf{Phase diagram:}}
There are three main phases that emerge for the networks. These can be characterized with just $m$ and $q_\text{EA}$, due to the fact $r$ and $r_d$ are not independent parameters, and $q$ is nonzero for all nonzero temperatures.  First is the paramagnetic phase found at high temperatures, characterized by $m=0$, $q_\text{EA}=0$. In this phase thermal fluctuations are so large that no order can form. At lower temperatures a spin glass phase may be found, characterized by $m=0$, $q_\text{EA}>0$. In this phase the network becomes trapped in spurious narrow free energy minima in which no Mattis magnetization is condensed. Of particular interest is the retrieval phase, characterized by $m>0$, $q_\text{EA}>0$. Within this phase the network is performing its intended function of retrieving an encoded pattern; the target Mattis magnetization $m$ is condensed.

Within the retrieval phase there are two distinct regions, called the global and local retrieval phases. Within the global retrieval phase, the retrieval solution of the equations of state is the global free energy minimum. Within the local retrieval phase, the free energy of the retrieval solution is greater than that of the spin glass solution, making the retrieval solution metastable. By comparing the retrieval and spin glass free energies, the boundary between these phases may be resolved.

The second-order spin glass transitions for the classical and quantum networks are found analytically by expanding the equations for $q_\text{EA}$ (and $q$ in the quantum case) to lowest order in $q_\text{EA}$ about the paramagnetic solution $(m=0,q_\text{EA}=0)$. For the QVHN, we have calculated the glass transition temperature $T_\text{sg,qu}(\alpha)$,
though the solution is cumbersome to write. When the system is minimally disordered with $\alpha\rightarrow 0$, we find $T_\text{sg,qu}(0)=1$. The transition temperature then increases monotonically with $\alpha$. For the CVHN, we find that $T_\text{sg,cl}(\alpha)=(1+\sqrt{\alpha/3})/3$. The quantum and rescaled classical glass transition curves are shown in Fig.~\ref{fig:quant_PD}.

In the limit $\alpha\rightarrow 0$, the pattern retrieval transitions for both networks are second-order. Independent of the static approximation, Eq.~(\ref{eq:q_eqs_m}) for the QVHN becomes $m=\tanh(\beta m)$, which gives us the retrieval transition temperature $T_\text{r,qu}(0)=1$. In the same limit, we find that Eq.~(\ref{eq:cl_eqs_m}) for the CVHN becomes $m=\coth(\beta m)-(\beta m)^{-1}$, which leads to $T_\text{r,cl}(0)=1/3$. Resolving the retrieval transitions for $\alpha>0$ requires numerically solving the equations of state.

\begin{figure}[t]
  \centering
  \subfloat{\includegraphics[width=\linewidth]{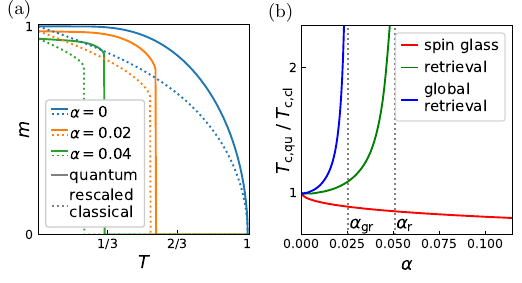}\label{fig:m_curves}}
  \subfloat{\label{fig:ratios}}
  \caption{(a) The Mattis magnetization $m$ at several values of pattern loading $\alpha$ for the quantum, classical, and rescaled classical networks. The quantum network displays a less steep initial decrease in $m$ as $T$ increases compared to the classical networks. At $T=0$ the Mattis magnetizations of all three networks coincide. (b) The ratios of the quantum critical temperatures to the classical critical temperatures for the spin glass, retrieval, and global retrieval transitions. The global retrieval ratio appears to diverge at $\alpha_\text{gr}\approx 0.0252$ and the retrieval ratio appears to diverge at $\alpha_\text{r}\approx 0.0508$. The stabilizing quantum effects are greatest, relative to the classical network, as the networks approach capacity.}
  \label{fig:replica_curves}
\end{figure}

We note that, when $\alpha\rightarrow 0$, $T_\text{sg,qu}=T_\text{r,qu}=3T_\text{sg,cl}=3T_\text{r,cl}$. This difference of a factor of three is fully explained by the difference in susceptibility between a spin-1/2 and a corresponding classical spin. Indeed, if $\vec\sigma_\text{qu}$ is a vector of Pauli matrices and $\vec\sigma_\text{cl}$ is a classical vector of unit magnitude, then $|\vec\sigma_\text{qu}|^2=3|\vec\sigma_\text{cl}|^2$. The fluctuation-dissipation theorem~\cite{Kubo1957-or} then tells us that the high-temperature susceptibility of a spin-1/2 is three times larger than that of a classical spin of the same magnitude, leading to a three times higher transition temperature for $\alpha \to 0$. This inherent difference can be offset by rescaling the spins of the classical network to have a magnitude of $\sqrt 3$, as we have done in Figs.~\ref{fig:quant_PD}, \ref{fig:basins} and \ref{fig:replica_curves}. Doing so allows us to reveal clearly nontrivial stability enhancement that arises for extensive pattern loading with $\alpha>0$.

\paragraph{\textbf{Numerical Results:}} The phase diagram that results from numerically solving the equations of state is shown in Fig.~\ref{fig:quant_PD}. We see that, compared to the CVHN, the QVHN exhibits higher retrieval transition temperatures for $0<\alpha<\alpha_\text{r}\approx0.0508$ and higher global retrieval transition temperatures for $0<\alpha<\alpha_\text{gr}\approx0.0252$, even after the aforementioned classical spin rescaling. This demonstrates that the quantum spin fluctuations have a stabilizing effect collectively, on top of the expected factor-of-three enhancement arising from the difference in single-spin susceptibilities.

This effect is further demonstrated by the comparison of the Mattis magnetization curves of the rescaled CVHN and QVHN in Fig.~\ref{fig:m_curves}. The rescaled CVHN shows a faster decay in retrieval temperature as $\alpha$ increases than the QVHN. Additionally, we see that the quantum Mattis magnetization curves stay near their zero temperature values up to higher temperatures before decaying.

The degree to which quantum effects stabilize retrieval relative to the rescaled CVHN is quantified by the ratios of the retrieval temperatures. These are plotted in Fig.~\ref{fig:ratios}. Of the three transition types, only $T_\text{sg,qu}/T_\text{sg,cl}$ decreases with increasing $\alpha$, signaling relative robsutness of the classical glass phase. The other two ratios increase with $\alpha$ up to its critical value. We observe that $T_\text{gr,qu}/T_\text{gr,cl}$ appears to diverge as $\alpha$ approaches $\alpha_\text{gr}$. Similarly, $T_\text{r,qu}/T_\text{r,cl}$ appears to diverge as $\alpha$ approaches $\alpha_\text{r}$. This means that the quantum stabilizing effects are greatest as the networks approach capacity.

As an additional verification of our results for the CVHN, we numerically simulated instances of the network and estimated their equilibrium behavior using a Gibbs sampling procedure~\cite{geman_stochastic_1984, miyatake_implementation_1986}. Details of the simulation and sampling procedure are found in the Supplemental Material~\cite{Sup}.

\begin{figure}[t]
  \centering
  \subfloat{\includegraphics[width=\linewidth]{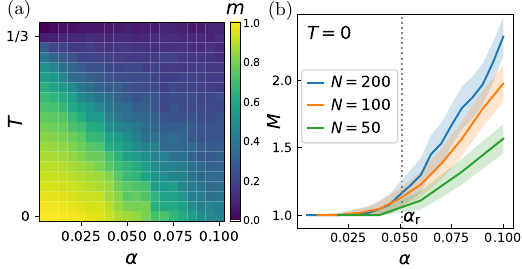}\label{fig:simulated_mattis}}
  \subfloat{\label{fig:mixtures}}
  \caption{(a) The average equilibrium Mattis magnetization $m$ of a state along its initial pattern in a simulated CVHN with $N=200$ spins,  averaged over 40 network instances and $5$ random initializations each. High average $m$ (indicating retrieval) is seen in the region predicted by the replica analysis, though the transition in $\alpha$ at fixed temperature is seen to be smoother. (b) The average number $M$ of condensed Mattis magnetizations for the simulated equilibrium states at $T=0$ as a function of pattern loading $\alpha$, averaged over 1000 network instances and $5$ random initializations each. $M$ begins to increase near the value predicted by the replica analysis for the retrieval transition.}
  \label{fig:simulation_results}
\end{figure}

The average target Mattis magnetization at various temperatures and pattern loadings is shown in Fig.~\ref{fig:simulated_mattis}. We see high $m$ (indicating retrieval) in roughly the region predicted by the replica analysis. In particular, we observe $m$ drop to zero around $T=1/3$ for $\alpha\to 0$, as expected. However, while $m$ decreases with increasing $\alpha$, the decrease is gentler than the phase transition predicted by the replica analysis. We surmise that this is because the replica treatment only considers the case with, at most, one condensed Mattis magnetization. But, in the simulated networks, it is possible for the final state to be a mixture state with significant overlap with multiple patterns. To test this, at $T=0$ we sorted all the Mattis magnetizations of each final state in decreasing order, counted the number of magnetizations above the largest break in the sequence, and averaged over network instances and initializations. The results are shown in Fig.~\ref{fig:mixtures}. We observe an increase in the thus defined number of condensed magnetizations near the critical $\alpha$ value predicted from the replica analysis. The appearance of these states with multiple condensed Mattis magnetizations may explain the slow decrease of $m$ with $\alpha$ seen in Fig.~\ref{fig:simulated_mattis}. 

\paragraph{\textbf{Interpretation:}} The enhancement in retrieval stability due to quantum effects can be interpreted analogously to the phenomenon of quantum order-by-disorder~\cite{Villain1980-vk,Shender1982,Henley1989-qj}. Within this picture, we can view the quantum free energy landscape as a rough classical landscape smoothed by quantum fluctuations (see Fig.~\ref{fig:smooth}). The free energies of the minima will be lifted by quantum fluctuations above the corresponding classical minima proportionally to the local curvature. One expects that the classical spin glass solutions reside in deep rugged local minima in the high-dimensional phase space of the model,  while the retrieval solutions are surrounded by a larger basin of attraction with lower curvature. Therefore, the spin glass solution is lifted to higher free energies by quantum fluctuations than the memory states, relatively stabilizing the global retrieval phase.  This view is supported by results in Fig.~\ref{fig:basins}, which show the retrieval solution occupying a broader basin in order parameter space than the spin glass. We also find that the quantum stabilization effects increase (enhanced ratio of $T_{\text{r,qu}} / T_{\text{r,cl}}$) as pattern loading $\alpha$ is increased, which we anticipate increases the roughness of the classical free energy landscape and concomitantly, renormalization by quantum fluctuations.  


Besides an increase in the retrieval transition temperature, we see that quantum fluctuations also enhance the accuracy of stored memory patterns, particularly as temperature is increased; see Fig.~\ref{fig:m_curves}. 
We speculate that the classical free energy landscape in spin configuration space exhibits many minima that are slightly disparate from the intended memory patterns. Quantum fluctuations help smooth out these irregularities by merging such minima, yielding one closer to the intended memory. However, purely quantum tunneling may not be strong enough to merge energetically disparate minima at zero temperature; thus increasing temperature likely aids the observed relative enhancement, at least for smaller temperatures. 
%

Overall, these results demonstrate that quantum effects can be used to stabilize pattern retrieval in an associative memory. In the quantum vector Hopfield network, there is a base enhancement due to the difference in single-spin susceptibilities between classical and quantum spins. But, beyond that, there is a nontrivial enhancement arising from collective quantum fluctuations of the spins. This relative enhancement increases with pattern loading up to capacity.

\paragraph{\textbf{Acknowledgments:}}
We thank Maxim Khodas for useful discussions and helpful comments on the manusript. R.D.B., S.B., K.A. and I.M. acknowledge support by the US Department of Energy, Office of Science, Basic Energy Sciences, Materials Sciences and Engineering Division. 

\bibliography{paper_bib}

\onecolumngrid
\clearpage


\section*{Supplementary material}

# Supplemental Material for "Quantum-stabilized pattern retrieval: the quantum vector Hopfield network"

Richard D. Barney,[1] Sharba Bhattacharjee,[1] Victor Galitski,[2] Kartiek Agarwal,[1] and Ivar Martin[1]

[1]*Materials Science Division, Argonne National Laboratory, Argonne, Illinois 60439, USA*

[2]*Joint Quantum Institute, Department of Physics, University of Maryland, College Park, Maryland 20742, USA*

## I. REPLICA DERIVATION OF THE FREE ENERGY DENSITY

We consider the replicated partition function $Z^n$ with $n$ replicas. We make a Suzuki-Trotter transformation [S1] by inserting spin coherent states [S2, S3] $|\vec{g}_i^\gamma(\tau)\rangle$ at each imaginary time step, with $\gamma$ being the replica index. That is, we multiply by the identity

$$1 \propto \int [d|\vec{g}(\tau)\rangle] \left( \prod_{i\gamma} |\vec{g}_i^\gamma(\tau)\rangle \right) \left( \prod_{i\gamma} \langle \vec{g}_i^\gamma(\tau)| \right) \tag{S1}$$

at each imaginary time step. In the continuous time limit, the replicated partition function is

$$Z^n = \int \mathscr{D}\vec{g} \exp\left[ -\int d\tau \sum_{i\gamma} \langle \partial_\tau \vec{g}_i^\gamma | \vec{g}_i^\gamma \rangle + N^{-1} \int d\tau \sum_{\mu,\gamma,i<j} (\vec{\xi}_i^\mu \cdot \vec{g}_i^\gamma)(\vec{\xi}_j^\mu \cdot \vec{g}_j^\gamma) \right], \tag{S2}$$

where $\mathscr{D}\vec{g} = \prod_{i,\gamma,\tau} d\vec{g}_i^\gamma(\tau)\delta(|\vec{g}_i^\gamma(\tau)|^2 - 1)$. For compactness, we refrain from writing the explicit $\tau$ dependence within terms which are local in imaginary time. Integrals over imaginary time are on the interval $(0, \beta)$.

We define the overlaps

$$m^{\mu\gamma}(\tau) = N^{-1} \sum_i \vec{\xi}_i^\mu \cdot \vec{g}_i^\gamma(\tau) \tag{S3}$$

and rewrite the replicated partition function as

$$Z^n = \int \mathscr{D}\vec{g}\mathscr{D}m \exp\left[ -\int d\tau \sum_{i\gamma} \langle \partial_\tau \vec{g}_i^\gamma | \vec{g}_i^\gamma \rangle + \int d\tau \sum_{\gamma\mu} \left( -\frac{N}{2}(m^{\mu\gamma})^2 + m^{\mu\gamma} \sum_i \vec{\xi}_i^\mu \cdot \vec{g}_i^\gamma - \frac{1}{2N} \sum_i (\vec{\xi}_i^\mu \cdot \vec{g}_i^\gamma)^2 \right) \right]. \tag{S4}$$

We now wish to average over the uncondensed patterns. We can see that, in order for the energy to be extensive, $O(N)$ of the overlaps must be $O(N^{-1/2})$. So, for an uncondensed $\vec{\xi}_i^\mu$,

$$\begin{aligned}
&\overline{\exp\left[ \vec{\xi}_i^\mu \cdot \int d\tau \sum_\gamma m^{\mu\gamma} \vec{g}_i^\gamma - \frac{1}{2N} \int d\tau \sum_\gamma (\vec{\xi}_i^\mu \cdot \vec{g}_i^\gamma)^2 \right]} \\
&\qquad = 1 + \frac{1}{2} \left( \overline{\left[ \vec{\xi}_i^\mu \cdot \left( \int d\tau \sum_\gamma m^{\mu\gamma} \vec{g}_i^\gamma \right) \right]^2} - N^{-1} \int d\tau \sum_\gamma \overline{(\vec{\xi}_i^\mu \cdot \vec{g}_i^\gamma)^2} \right) + O(N^{-2}) \\
&\qquad = \exp\left[ \frac{1}{6} \left( \left| \int d\tau \sum_\gamma m^{\mu\gamma} \vec{g}_i^\gamma \right|^2 - N^{-1} \int d\tau \sum_\gamma |\vec{g}_i^\gamma|^2 \right) + O(N^{-2}) \right]
\end{aligned}. \tag{S5}$$

We specialize to the case of a single possibly condensed overlap $m$ and, without loss of generality, let it correspond to

the pattern $\vec{\xi}^1$. We may then use the equation above and integrate out all the uncondensed overlaps to find

$$\overline{\int \prod_{\mu>1,\gamma} \mathcal{D}m^{\mu\gamma} \exp \int d\tau \sum_{\mu>1,\gamma} \left( -\frac{N}{2}(m^{\mu\gamma})^2 + m^{\mu\gamma} \sum_i \vec{\xi}_i^\mu \cdot \vec{g}_i^\gamma - \frac{1}{2N} \sum_i (\vec{\xi}_i^\mu \cdot \vec{g}_i^\gamma)^2 \right)} \tag{S6}$$

$$= \exp\left( -\frac{\alpha}{6} \int d\tau \sum_{\gamma,i} |\vec{g}_i^\gamma|^2 \right) \int \prod_{\mu>1,\gamma} \mathcal{D}m^{\mu\gamma} \exp\left[ -\frac{N}{2} \int d\tau \sum_\gamma \sum_{\mu>1} (m^{\mu\gamma})^2 + \frac{1}{6} \sum_{\mu>1} \sum_i \left| \int d\tau \sum_\gamma m^{\mu\gamma} \vec{g}_i^\gamma \right|^2 \right]$$
$$= \exp\left( -\frac{\alpha}{6} \int d\tau \sum_{\gamma,i} |\vec{g}_i^\gamma|^2 - \frac{p}{2} \mathrm{Tr} \ln A \right), \tag{S7}$$

$$A^{\gamma\gamma'}(\tau,\tau') = \delta^{\gamma\gamma'} \delta(\tau-\tau') - \frac{1}{3N} \sum_i \vec{g}_i^\gamma(\tau) \cdot \vec{g}_i^{\gamma'}(\tau'). \tag{S8}$$

We now define the configuration overlap parameters

$$q^{\gamma\gamma'}(\tau,\tau') = N^{-1} \sum_i \vec{g}_i^\gamma(\tau) \cdot \vec{g}_i^{\gamma'}(\tau') \tag{S9}$$

and enforce them with corresponding Lagrange multipliers $r^{\gamma\gamma'}(\tau,\tau')$. That is, we multiply the replicated partition function by the identity

$$1 \propto \int \mathcal{D}q\mathcal{D}r \exp\left[ -\frac{p}{2} \int d\tau d\tau' \sum_{\gamma\gamma'} r^{\gamma\gamma'}(\tau,\tau') \left( q^{\gamma\gamma'}(\tau,\tau') - N^{-1} \sum_i \vec{g}_i^\gamma(\tau) \cdot \vec{g}_i^{\gamma'}(\tau') \right) \right]. \tag{S10}$$

We make the replica symmetric ansatz and use the fact that the system is time translation invariant. This means that

$$q^{\gamma\gamma'}(\tau,\tau') = \begin{cases} q_0(\tau-\tau'), & \gamma=\gamma' \\ q_{\mathrm{EA}}, & \gamma\neq\gamma' \end{cases}, \quad r^{\gamma\gamma'}(\tau,\tau') = \begin{cases} r_0(\tau-\tau'), & \gamma=\gamma' \\ r_d, & \gamma\neq\gamma' \end{cases}. \tag{S11}$$

Under this ansatatz, $A$ is a circulant matrix in replica space, and the diagonal elements are circulant in time. Therefore, $A$ may be diagonalized by means of a discrete Fourier transform in replica space followed by a continuous Fourier transform in time. Doing so, we find

$$\mathrm{Tr} \ln A = n \left( \ln\left(1 - \beta(q_0 - q_{\mathrm{EA}})/3\right) - \frac{\beta q_{\mathrm{EA}}/3}{1 - \beta(q_0 - q_{\mathrm{EA}})/3} + \sum_{b\neq 0} \ln(1 - \tilde{q}_b^0/3) \right) + O(n^2), \tag{S12}$$

where $\tilde{q}_b^0 = \int d\tau e^{i\omega_b \tau} q_0(\tau)$ is the Fourier transform of $q_0(\tau)$ at Matsubara frequency $\omega_b = \frac{2\pi b}{\beta}$.

Keeping only $O(Nn)$ terms in the action, we find that the disorder averaged replicated partition function is

$$\overline{Z^n} = \int dr_d dq_{\mathrm{EA}} \mathcal{D}r_0 \mathcal{D}q_0 \exp\left[ -\frac{pn}{2} \left( \ln\left(1 - \beta(q_0 - q_{\mathrm{EA}})/3\right) - \frac{\beta q_{\mathrm{EA}}/3}{1 - \beta(q_0 - q_{\mathrm{EA}})/3} + \sum_{b\neq 0} \ln(1 - \tilde{q}_b^0/3) \right) \right.$$
$$\left. - \beta \frac{pn}{6} q_0(0) + \beta^2 \frac{pn}{6} \left( r_d q_{\mathrm{EA}} - \int \frac{d\tau}{\beta} r_0 q_0 \right) \right] \overline{F}, \tag{S13}$$

$$F = \int \mathcal{D}\vec{g}\mathcal{D}m \exp\Bigg[-\int d\tau \sum_{i\gamma} \langle \partial_\tau \vec{g}_i^\gamma | \vec{g}_i^\gamma \rangle - \frac{N}{2}\int d\tau \sum_{\gamma\mu} (m^{\mu\gamma})^2 + \frac{\alpha}{6} r_d \int d\tau d\tau' \sum_{\gamma\neq\gamma',i} \vec{g}_i^\gamma(\tau)\cdot \vec{g}_i^\gamma(\tau') \tag{S14}$$

$$+\frac{\alpha}{6}\int d\tau d\tau' r_0(\tau-\tau') \sum_{\gamma i} \vec{g}_i^\gamma(\tau)\cdot\vec{g}_i^\gamma(\tau') + \int d\tau \sum_{\gamma\mu} m^{\mu\gamma}\sum_i \vec{\xi}_i^\mu \cdot \vec{g}_i^\gamma\Bigg]$$

$$= \Bigg\{\int \mathcal{D}\vec{g}\mathcal{D}m \exp\Bigg[-\int d\tau \sum_\gamma \langle \partial_\tau \vec{g}^\gamma | \vec{g}^\gamma\rangle - \frac{1}{2}\int d\tau \sum_{\gamma\mu}(m^{\mu\gamma})^2 + \int d\tau \sum_{\gamma\mu} m^{\mu\gamma}\vec{\xi}^\mu \cdot \vec{g}^\gamma \tag{S15}$$

$$+\frac{\alpha}{6} r_d \int d\tau d\tau' \sum_{\gamma\gamma'} \vec{g}^\gamma(\tau)\cdot\vec{g}^\gamma(\tau') + \frac{\alpha}{6}\int d\tau d\tau' (r_0(\tau-\tau') - r_d)\sum_\gamma \vec{g}^\gamma(\tau)\cdot\vec{g}^\gamma(\tau')\Bigg]\Bigg\}^N$$

$$= \Bigg\{\int d\!\!\!/\vec{h}\Bigg[\int \mathcal{D}\vec{g}\mathcal{D}m \exp\Bigg(-\int d\tau \langle\partial_\tau \vec{g}|\vec{g}\rangle - \frac{1}{2}\int d\tau m^2 + \int d\tau m \vec{\xi}^1\cdot\vec{g} \tag{S16}$$

$$+\sqrt{\frac{\alpha r_d}{3}}\vec{h}\cdot\int d\tau \vec{g} + \frac{\alpha}{6}\int d\tau d\tau' (r_0(\tau-\tau')-r_d)\vec{g}(\tau)\cdot\vec{g}(\tau')\Bigg)\Bigg]^n\Bigg\}^N. \tag{S17}$$

In the last equality we have performed a Hubbard-Stratonovich transformation and introduced the time independent field $\vec{h}$ with the measure $d\!\!\!/\vec{h} = dh_x dh_y dh_z (2\pi)^{-3/2} e^{-|\vec{h}|^2/2}$.

In order to decouple the remaining quadratic interaction with an imaginary time-independent Hubbard-Stratonovich field, we make the static approximation, which is that all order parameters are time independent. We may then write

$$F = \int dm e^{-\beta N n m/2}\left\{\int d\!\!\!/\vec{h}\left[\int \mathcal{D}\vec{g} d\!\!\!/\vec{\ell} \exp\left(-\int d\tau \langle\partial_\tau \vec{g}|\vec{g}\rangle + \vec{B}\cdot\int d\tau \vec{g}\right)\right]^n\right\}^N, \tag{S18}$$

$$\vec{B} = m\vec{\xi}^1 + \sqrt{\frac{\alpha r_d}{3}}\vec{h} + \sqrt{\frac{\alpha r}{3}}\vec{\ell}. \tag{S19}$$

Using this, and Eq. (S13), we obtain the free energy density $f$ through the replica trick [S4, S5]

$$f = \lim_{n\to 0}\frac{1-\overline{Z^n}}{\beta N n}. \tag{S20}$$

Doing so, and making the change of variables $q_0 - q_{\mathrm{EA}} \to q$, $r_0 - r_d \to r$, yields Eqs. (5) and (6) in the main text.

## II. MEAN-FIELD DERIVATION OF THE EQUATIONS OF STATE

As in the replica calculation of the previous section, the first step is performing a Suzuki-Trotter transformation [S1], inserting spin coherent states [S2, S3] $|\vec{g}_i(\tau)\rangle$ at each imaginary time step. In the continuous time limit, we obtain the effective single-site partition function

$$Z_i = \int \mathcal{D}m\mathcal{D}\vec{g} \exp\int_0^\beta d\tau \left(-\langle\partial_\tau \vec{g}_i(\tau)|\vec{g}_i(\tau)\rangle + \sum_\mu m^\mu(\tau)\vec{\xi}_i^\mu\cdot\vec{g}_i(\tau)\right), \tag{S21}$$

$$m^\mu(\tau) = N^{-1}\sum_i \vec{\xi}_i^\mu\cdot\vec{g}_i(\tau), \tag{S22}$$

where $\mathcal{D}\vec{g} = \prod_{i,\tau} d\vec{g}_i(\tau)\delta(|\vec{g}_i(\tau)|^2-1)$, and the $m^\mu(\tau)$ are the Mattis magnetizations. Technically, this introduces self-interaction terms to the action, but their effect is negligible in the thermodynamic limit.

In order to integrate over the $\vec{g}_i(\tau)$, we make the static approximation, i.e. that the overlaps $m^\mu(\tau)$ are imaginary time-independent. The integral over $\vec{g}$ is then just the partition function of a spin in a constant field. This yields the single-site partition function

$$Z_i \propto \int Dm \cosh(\beta|\vec{B}_i|), \quad \vec{B}_i = \sum_\mu m^\mu \vec{\xi}_i^\mu. \tag{S23}$$

Since $\vec{B}_i$ is a sum of an extensive number of random variables, and $m^\mu$ with $\mu > 1$ is weakly dependent on the $\vec{\xi}_i^\mu$, we may apply the central limit theorem to treat $\vec{B}_i$ as a Gaussian random variable. We assume that there is, at most, one condensed overlap $m$ and let it, without loss of generality, correspond to the pattern $\vec{\xi}^1$. We can then write

$$\begin{aligned}\vec{B}_i &= m\vec{\xi}_i^1 + \sum_{\mu>1}\langle m^\mu\rangle\vec{\xi}_i^\mu + \sum_{\mu>1}(m^\mu - \langle m^\mu\rangle)\vec{\xi}_i^\mu \\ &= m\vec{\xi}_i^1 + \sqrt{\frac{\alpha r_d}{3}}\vec{h}_i + \sqrt{\frac{\alpha r}{3}}\vec{\ell}_i,\end{aligned} \tag{S24}$$

$$\alpha r_d = \sum_{\mu>1}\overline{\langle m^\mu\rangle^2}, \tag{S25}$$

$$\alpha r = \sum_{\mu>1}\left(\overline{\langle(m^\mu)^2\rangle} - \overline{\langle m^\mu\rangle^2}\right), \tag{S26}$$

where $\vec{h}_i$ and $\vec{\ell}_i$ are Gaussian random variables with 0 mean and unit variance element-wise. The integrals over the $m^\mu$ in the single-site partition function $Z_i$ are then reduced to a single integral over $\vec{\ell}_i$ by the central limit theorem. Note that this makes $Z_i$ independent of the direction of $\vec{\xi}_i^1$.

After performing the integral over $\vec{\ell}_i$ we obtain the effective single-site free energy

$$f_i = -\beta^{-1}\ln Z_i \tag{S27}$$

$$= -\frac{\beta\alpha}{6}r - w(|\vec{b}_i|), \tag{S28}$$

$$w_{\text{qu}}(v) = \beta^{-1}\ln\left(\cosh(\beta v) + \frac{\beta^2\alpha r}{3}\text{sinhc}(\beta v)\right), \tag{S29}$$

$$\vec{b}_i = m\vec{\xi}_i^1 + \sqrt{\frac{\alpha r_d}{3}}\vec{h}_i, \tag{S30}$$

where $\text{sinhc}(x) = \sinh(x)/x$. This gives us the average effective single-site free energy

$$f_{\text{ss}} = N^{-1}\sum_i f_i \tag{S31}$$

$$= -\frac{\beta\alpha}{6}r - N^{-1}\sum_i w(|\vec{b}_i|) \tag{S32}$$

$$= -\frac{\beta\alpha}{6}r - \int d\!\!\!/\vec{h}\, w(|\vec{b}|) \tag{S33}$$

$$= -\frac{\beta\alpha}{6}r - m^{-1}\int d\!\!\!/\rho\, v w(v), \tag{S34}$$

$$v = m + \sqrt{\frac{\alpha r_d}{3}}\rho, \tag{S35}$$

where $d\!\!\!/\vec{h} = dh_x dh_y dh_z (2\pi)^{-3/2}e^{-|\vec{h}|^2/2}$ and $d\!\!\!/\rho = d\rho(2\pi)^{-1/2}e^{-\rho^2/2}$. In reducing the integral over $\vec{h}$ to a one-dimensional integral, we have used the fact that, for any even function $\gamma(|\vec{b}|)$,

$$\int d\!\!\!/\vec{h}\,\gamma(|\vec{b}|) = m^{-1}\int d\!\!\!/\rho\, v\gamma(v), \tag{S36}$$

which can be proven by switching to spherical coordinates and integrating over the angular coordinates. Note that Eq. (S34) matches Eq. (8) in the main text.

Using the definition of the overlaps in Eq. (S22) along with Eqs. (S21), (S27), and (S31), we see that

$$\langle m^\mu\rangle = \frac{1}{\beta N}\partial_{m^\mu}\sum_i \ln Z_i = -\partial_{m^\mu} f_{\text{ss}}. \tag{S37}$$

For $\mu = 1$, this immediately leads to the equation for the condensed Mattis magnetization $m$, Eq. (10) in the main text.

We must also find $\langle m^\mu \rangle$ for $\mu > 1$ in order to calculate $r_d$ and $r$. Expanding Eq. (S37) to first order in $\langle m^\mu \rangle$ and solving for $\langle m^\mu \rangle$, we find

$$\langle m^\mu \rangle = -\frac{\partial_{m^\mu} f_{\rm ss}}{1 + \partial^2_{m^\mu} f_{\rm ss}}. \tag{S38}$$

We may write

$$\vec{b}_i = m\vec{\xi}^1_i + m^\mu \vec{\xi}^\mu_i + \sqrt{\frac{\alpha r_d}{3}}\vec{h}, \tag{S39}$$

without changing the distribution of $\vec{h}_i$ in the thermodynamic limit. This means that we can use Eq. (S39) within $f_{\rm ss}$ when we need to take a derivative with respect to $m^\mu$ and let $m^\mu \to 0$ otherwise. We note that

$$\partial_{m^\mu} f_i = \frac{d|\vec{b}_i|}{dm^\mu}\partial_{|\vec{b}_i|} f_i = -\frac{\vec{\xi}^\mu_i \cdot \vec{b}_i}{|\vec{b}_i|}\partial_{|\vec{b}_i|} w(|\vec{b}_i|), \tag{S40}$$

$$\partial^2_{m^\mu} f_i = -\partial_{m^\mu} \frac{\vec{\xi}^\mu_i \cdot \vec{b}_i}{|\vec{b}_i|}\partial_{|\vec{b}_i|} w(|\vec{b}_i|) = \left(\frac{(\vec{\xi}^\mu_i \cdot \vec{b}_i)^2}{|\vec{b}_i|^3} - |\vec{b}_i|^{-1}\right)\partial_{|\vec{b}_i|} w(|\vec{b}_i|) - \frac{(\vec{\xi}^\mu_i \cdot \vec{b}_i)^2}{|\vec{b}_i|^2}\partial^2_{|\vec{b}_i|} w(|\vec{b}_i|). \tag{S41}$$

So

$$\begin{aligned}\partial^2_{m^\mu} f_{\rm ss} &= N^{-1}\sum_i \partial^2_{m^\mu} f_i = -\frac{1}{3}\int d\!\!\!/\vec{h}\left(2|\vec{b}|^{-1}\partial_{|\vec{b}|} w(|\vec{b}|) + \partial^2_{|\vec{b}|} w(|\vec{b}|)\right) \\ &= (m\alpha r_d)^{-1}\int d\!\!\!/\rho(1-\rho^2) v w(v)\end{aligned} \tag{S42}$$

and

$$\begin{aligned}(\partial_{m^\mu} f_{\rm ss})^2 &= N^{-2}\sum_{ij}(\partial_{m^\mu} f_i)(\partial_{m^\mu} f_j) = \frac{1}{3N}\int d\!\!\!/\vec{h}\left(\partial_{|\vec{b}|} w(|\vec{b}|)\right)^2 \\ &= \frac{1}{3N}m^{-1}\int d\!\!\!/\rho v\left(\partial_v w(v)\right)^2.\end{aligned} \tag{S43}$$

In deriving these relations we have used Eq. (S36) and integration by parts. Using Eqs. (S38) and (S42)-(S43) in Eq. (S25), we find Eq. (13) in the main text for $r_d$, along with Eqs. (11) and (12) for $q$ and $q_{\rm EA}$ respectively.

We now calculate $\langle (m^\mu)^2 \rangle_c$ in order to determine $r$. From Eq. (S37) we see that

$$\langle (m^\mu)^2 \rangle_c = (\beta N)^{-2}\partial^2_{m^\mu}\sum_i \ln Z_i = -(\beta N)^{-1}\partial^2_{m^\mu} f_{\rm ss} = (\beta N)^{-1}\partial_{m^\mu}\langle m^\mu \rangle. \tag{S44}$$

Using this and Eq. (S38) in Eq. (S26), neglecting $\partial^3_{m^\mu} f$ since $m^\mu$ is well-approximated as a Gaussian random variable, we find Eq. (13) in the main text for $r$.

With a different definition of $w(v)$, these equations also hold for the classical network. Taking the spin number to $\infty$ in Eq. (S21) restricts $\vec{g}(\tau)$ to imaginary time-independent paths, yielding the single-site free energy

$$f_i = -\frac{\beta\alpha}{6}r - \beta^{-1}\ln \operatorname{sinhc}(\beta|\vec{b}_i|). \tag{S45}$$

By comparison to Eq. (S28), we see that, in the classical case,

$$w_{\rm cl}(v) = \beta^{-1}\ln \operatorname{sinhc}(\beta v). \tag{S46}$$

Inserting this in the equations of state, and using the fact that $q + q_{\rm EA} = 1$ to eliminate the equations for $r$ and $q$, the classical equations of state are obtained, Eqs. (14)-(16) in the main text.

## III. RELAXING THE STATIC APPROXIMATION

If we do not make the static approximation, we obtain the free energy density

$$f = \frac{\alpha}{2}\left[\beta^{-1}\sum_{b=-\infty}^{\infty}\left(\ln(1-\tilde{q}_b/3)+\frac{\tilde{q}_b}{3}\right)-\frac{q_{\rm EA}\tilde{q}_0^c/9}{1-\tilde{q}_0^c/3}+\frac{1}{3}\left(\beta\sum_{b=-\infty}^{\infty}\tilde{r}_b\tilde{q}_b+q_{\rm EA}\tilde{r}_0^c+\tilde{q}_0^c r_d\right)\right]+\frac{1}{2}m^2+f_{\rm ss} \tag{S47}$$

$$f_{\rm ss} = -\beta^{-1}\ln\int \mathscr{D}\vec{h}\mathscr{D}\vec{g}\exp\left[-\int d\tau\langle\partial_\tau\vec{g}|\vec{g}\rangle+\left(m\vec{\xi}^1+\sqrt{\frac{\alpha r_d}{3}}\vec{h}\right)\cdot\tilde{\vec{g}}_0+\frac{\alpha}{6\beta}\sum_{b=-\infty}^{\infty}\tilde{r}_b\tilde{\vec{g}}_{-b}\cdot\tilde{\vec{g}}_b\right], \tag{S48}$$

where $\tilde{r}_b$, $\tilde{q}_b$, and $\tilde{\vec{g}}_b$ are Fourier components. We expand $f_{\rm ss}$ to first order in the $\tilde{r}_b$ for $b\neq 0$, yielding

$$f_{\rm ss} \approx -\beta^{-1}\int \mathscr{D}\vec{h}\ln Z_0-\frac{\alpha}{6\beta^2}\sum_{b\neq 0}\int \mathscr{D}\vec{h}\langle\tilde{\vec{g}}_{-b}\cdot\tilde{\vec{g}}_b\rangle_0\tilde{r}_b \tag{S49}$$

where $Z_0$ is the static effective single-site partition function and $\langle\cdots\rangle_0$ denotes the thermal expectation with respect to $Z_0$.

Inserting Eq. (S49) into Eq. (S47) and varying with respect to $\tilde{q}_b$ and $\tilde{r}_b$ with $b\neq 0$, we find the saddle point equations

$$\tilde{q}_b = \frac{\tilde{r}_b}{1+\tilde{r}_b}, \tag{S50}$$

$$\tilde{q}_b = \beta^{-1}\int \mathscr{D}\vec{h}\langle\tilde{\vec{g}}_{-b}\cdot\tilde{\vec{g}}_b\rangle_0 \tag{S51}$$

We see that when $\tilde{r}_b$ is small, $\tilde{q}_b$ is small also. We may then expand the free energy density to second order in $\tilde{q}_b$, which is the first nonvanishing correction. We find that the lowest-order correction to the free energy density arising from quantum fluctuations of the order parameters is

$$\Delta f = -\frac{\alpha}{36\beta}\sum_{b\neq 0}\tilde{q}_b^2. \tag{S52}$$

By performing a Hubbard-Stratonovich transformation to decouple the remaining interaction term in $Z_0$ and using the result for the partition function of a spin-1/2 in a constant field, we find that

$$Z_0 = \int \mathscr{D}\vec{g}\exp\left[-\int d\tau\langle\partial_\tau\vec{g}|\vec{g}\rangle+\left(m\vec{\xi}^1+\sqrt{\frac{\alpha r_d}{3}}\vec{h}\right)\cdot\tilde{\vec{g}}_0+\frac{\alpha}{6}r|\tilde{\vec{g}}_0|^2\right] \tag{S53}$$

$$= \int \mathscr{D}\vec{\ell}\int \mathscr{D}\vec{g}\exp\int d\tau\left(-\langle\partial_\tau\vec{g}|\vec{g}\rangle+\vec{B}\cdot\vec{g}\right) \tag{S54}$$

$$= 2e^{\beta^2\alpha r/6}\left(\cosh(\beta|\vec{b}|)+\frac{\beta^2\alpha r}{3}\text{sinhc}(\beta|\vec{b}|)\right), \tag{S55}$$

with

$$\vec{B} = m\vec{\xi}^1+\sqrt{\frac{\alpha r_d}{3}}\vec{h}+\sqrt{\frac{\alpha r}{3}}\vec{\ell}, \qquad \vec{b} = m\vec{\xi}^1+\sqrt{\frac{\alpha r_d}{3}}\vec{h}. \tag{S56}$$

We then find that

$$\langle \tilde{\vec{g}}_{-b} \cdot \tilde{\vec{g}}_b \rangle_0 = \sum_a \int_0^\beta d\tau e^{-i\omega_b \tau} \langle \sigma_a(\tau)\sigma_a(0)\rangle_0 \tag{S57}$$

$$= Z_0^{-1} \int d\!\!\!{}^-\vec{\ell} \sum_{\zeta,\zeta'=\pm 1} e^{\beta\zeta|\vec{B}|} |\langle \zeta|\sigma_a|\zeta'\rangle|^2 \int_0^\beta d\tau e^{-i\omega_b\tau - (\zeta-\zeta')|\vec{B}|\tau} \tag{S58}$$

$$= Z_0^{-1} \int d\!\!\!{}^-\vec{\ell} \sum_a \sum_{\zeta,\zeta'=\pm 1} \frac{e^{\beta\zeta|\vec{B}|} - e^{\beta\zeta'|\vec{B}|}}{i\omega_b + (\zeta-\zeta')|\vec{B}|} |\langle \zeta|\sigma_a|\zeta'\rangle|^2 \tag{S59}$$

$$= \frac{16}{Z_0} \int d\!\!\!{}^-\vec{\ell} \frac{|\vec{B}| \sinh(\beta|\vec{B}|)}{\omega_b^2 + 4|\vec{B}|^2} \tag{S60}$$

$$< \frac{4}{Z_0} \int d\!\!\!{}^-\vec{\ell} |\vec{B}|^{-1} \sinh(\beta|\vec{B}|) \tag{S61}$$

$$= \frac{4}{Z_0} e^{\beta^2 \alpha r/6} |\vec{b}|^{-1} \sinh(\beta|\vec{b}|) \tag{S62}$$

$$< 2|\vec{b}|^{-1} \tanh(\beta|\vec{b}|). \tag{S63}$$

In Eq. (S58) we again introduced the Hubbard-Stratonovich field $\vec{\ell}$, allowing us to work with a system that is just a spin-1/2 in a constant field, then integrate over $\vec{\ell}$ afterward. This tells us that

$$\tilde{q}_b < \frac{2}{\beta} \int d\!\!\!{}^-\vec{h} |\vec{b}|^{-1} \tanh(\beta|\vec{b}|) \tag{S64}$$

$$= \frac{2}{\beta m} \int d\!\!\!{}^-\rho \tanh\left(\beta\left(m + \sqrt{\frac{\alpha r_d}{3}}\rho\right)\right) \tag{S65}$$

$$< \frac{2}{\beta m} \int d\!\!\!{}^-\rho \, \mathrm{sgn}\left(m + \sqrt{\frac{\alpha r_d}{3}}\rho\right) \tag{S66}$$

$$= \frac{2}{\beta m} \mathrm{erf}\left(\sqrt{\frac{3}{2\alpha r_d}} m\right) \tag{S67}$$

for $b \neq 0$, where $\mathrm{sgn}(x)$ is the sign function and $\mathrm{erf}(x)$ is the Gauss error function.

This means that, within the retrieval phase where $m \sim 1$, and noting that $|\mathrm{erf}(x)| < 1$,

$$|\Delta f| \lesssim \frac{\alpha}{36\beta^2}. \tag{S68}$$

Immediately we observe that this vanishes at zero temperature and when pattern loading is sub-extensive ($\alpha \to 0$). We compare this estimate of the correction to the free energy density with the static free energy densities $f_0$ obtained numerically for the retrieval phase. We find that $\Delta f/f_0 \lesssim 0.292\%$ in the worst case, with a substantial drop-off as we move away from the retrieval phase boundary. We conclude that the static approximation is appropriate for the retrieval phase of the QVHN.

## IV. SIMULATION OF THE CLASSICAL NETWORK

As an additional verification of our results for the CVHN, we numerically simulated instances of the network and estimated their equilibrium behavior at various temperatures and pattern loadings using a Gibbs sampling procedure [S6]. For each instance, a set of initial states was chosen, each with high initial overlap $m_0 \lesssim 1$ with one of the patterns. This was achieved by randomly choosing a set of patterns and then perturbing them by rotating each of their component spin vectors by the angle $\arccos(m_0)$ about a randomly chosen axis perpendicular to the spin vector. The local field acting on each spin $\vec{\sigma}_i$ is

$$\vec{b}_i = N^{-1} \sum_{j \neq i} \sum_\mu (\vec{\xi}_j^\mu \cdot \vec{\sigma}_j) \vec{\xi}_i^\mu. \tag{S69}$$

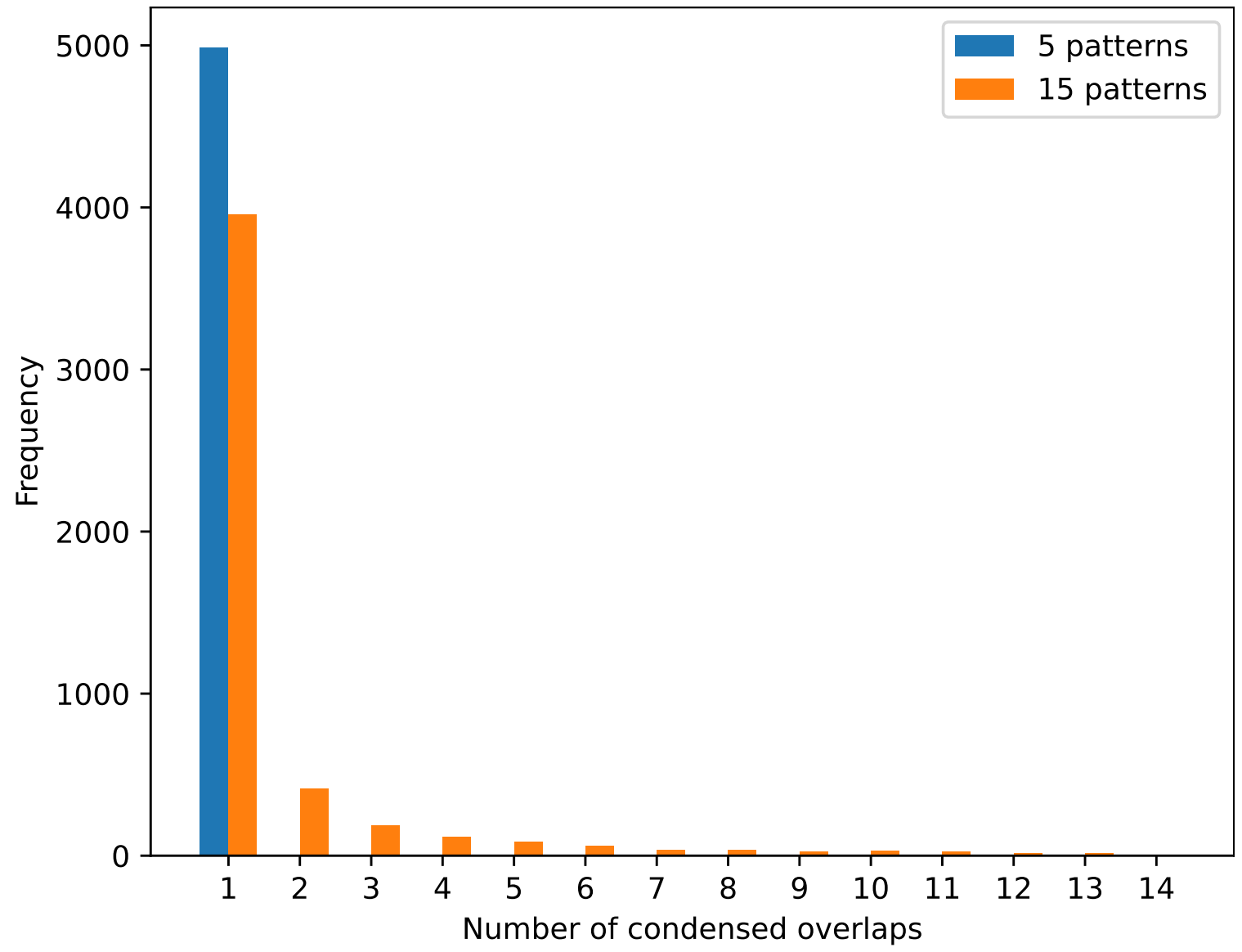


FIG. S1. Histograms of the number of condensed overlaps in simulated classical vector Hopfield networks with 200 spins at zero temperature, for 5 ($\alpha = 0.025$) and 15 patterns ($\alpha = 0.075$).

The spins were then updated asynchronously until the system approached equilibrium. At each step, a spin $\vec{\sigma}_i$ was chosen and rotated away from $\vec{b}_i$ about a random axis perpendicular to it by the random angle

$$\theta_i = \arccos\left[1 + \frac{T}{|\vec{b}_i|}\ln\left(1 - U_i\left(1 - e^{-2|\vec{b}_i|/T}\right)\right)\right], \tag{S70}$$

with $U_i \sim \text{Uniform}(0,1)$ [S7]. This ensured that the angle $\theta_i$ between the spin vector $\vec{\sigma}_i$ and its local field $\vec{b}_i$ had the probability density function

$$f(\cos(\theta_i)) = \frac{|\vec{b}_i| e^{|\vec{b}_i|\cos(\theta_i)/T}}{2T\sinh(|\vec{b}_i|/T)}, \tag{S71}$$

which is the expected Gibbs distribution for $\vec{\sigma}_i$ given the state of the rest of the system. The Mattis magnetizations were calculated for each instance and initialization after equilibrium was reached.

Our replica analysis only considers the states of the network which have significant overlap $m^\mu$ with at most one of the patterns (say $\xi^1$), i.e., $m^1 \sim 1$ in the retrieval phase but $m^\mu \sim 1/\sqrt{N}$ for $2 \le \mu \le p$ in all phases. However, in the simulated networks, it is possible for the model to reach a mixture state, where a finite number $M$ ($2 \le M \ll p$) of the patterns have high overlap with the state of the system. This may make the simulated results deviate from the phase transition predicted by the replica theoretic analysis. To estimate the prevalence of mixtures, for every simulated instance of the network, we obtained the quantity

$$M = \arg\max_\mu \left(|m^{(\mu)}| - |m^{(\mu+1)}|\right), \tag{S72}$$

where the Mattis magnetizations have been sorted in descending order by magnitude, $|m^{(1)}| \ge \ldots \ge |m^{(p)}|$. This identifies the largest gap between consecutive Mattis magnetizations and uses the number of patterns above that gap as an estimate of the number of condensed overlaps.

Fig. S1 shows the histograms of the number of condensed overlaps $M$ in simulated classical vector Hopfield networks with $N = 200$ spins at zero temperature, for $p = 5$ patterns and $p = 15$ patterns. The total number of samples is 5000 in each case (1000 randomly generated network instances and 5 random initializations for each network instance). We see that for fewer patterns ($p/N = 0.025$), $M = 1$ for almost all of the samples, indicating very low occurrence of mixture states. However, when the number of patterns is larger ($p/N = 0.075$), even though the state still has $M = 1$ in the majority of samples, there are a significant number of states with larger $M$, leading to an average $M$ of 1.67 in this case. In the main text we have shown the average $M$ at zero temperature as a function of $\alpha = p/N$, where the deviation of $M$ from 1 indicates the appearance of mixture states.



\end{document}